# Probing the Improbable: Methodological Challenges for Risks with Low Probabilities and High Stakes

Toby Ord, Rafaela Hillerbrand, Anders Sandberg[*]

Some risks have extremely high stakes. For example, a worldwide pandemic or asteroid impact could potentially kill more than a billion people. Comfortingly, scientific calculations often put very low probabilities on the occurrence of such catastrophes. In this paper, we argue that there are important new methodological problems which arise when assessing global catastrophic risks and we focus on a problem regarding probability estimation. When an expert provides a calculation of the probability of an outcome, they are really providing the probability of the outcome occurring, given that their argument is watertight. However, their argument may fail for a number of reasons such as a flaw in the underlying theory, a flaw in the modeling of the problem, or a mistake in the calculations. If the probability estimate given by an argument is dwarfed by the chance that the argument itself is flawed, then the estimate is suspect. We develop this idea formally, explaining how it differs from the related distinctions of model and parameter uncertainty. Using the risk estimates from the Large Hadron Collider as a test case, we show how serious the problem can be when it comes to catastrophic risks and how best to address it.

## 1. Introduction

Large asteroid impacts are highly unlikely events.[1] Nonetheless, governments spend large sums on assessing the associated risks. It is the high stakes that make these otherwise rare events worth examining. Assessing a risk involves consideration of both the stakes involved and the likelihood of the hazard occurring. If a risk threatens the lives of a great many people it is not only rational but morally imperative to examine the risk in some detail and to see what we can do to reduce it.

This paper focuses on low-probability high-stakes risks. In section 2, we show that the probability estimates in scientific analysis cannot be equated with the likelihood of these events occurring. Instead of the probability of the event occurring, scientific analysis gives the event's probability conditioned on the given argument being sound. Though this is the case in all probability estimates, we show how it becomes crucial when the estimated probabilities are smaller than a certain threshold.

To proceed, we need to know something about the reliability of the argument. To do so, risk analysis commonly falls back on the distinction between model and parameter uncertainty. We argue that this dichotomy is not well suited for incorporating information about the reliability of the theories involved in the risk assessment. Furthermore the distinction does not account for mistakes made unknowingly. In section 3, we therefore propose a three-fold distinction between an argument's theory, its model, and its calculations. While explaining this distinction in more detail, we illustrate it with historic examples of errors in each of the three areas. We indicate how specific risk assessment can make use of the proposed theory-model-calculation distinction in order to evaluate the reliability of the given argument and thus improve the reliability of their probability estimate for rare events.

---

[*] Future of Humanity Institute, University of Oxford.

[1] Experts estimate the annual probability as approximately one in a billion (Near-Earth Object Science Definition Team 2003).

Recently concerns have been raised that high-energy experiments in particle physics, such as the RHIC (Relativistic Heavy Ion Collider) at Brookhaven National Laboratory or the LHC (Large Hadron Collider) at CERN, Geneva, may threaten humanity. If these fears are justified, these experiments pose a risk to humanity that can be avoided by simply not turning on the experiment. In section 4, we use the methods of this paper to address the current debate on the safety of experiments within particle physics. We evaluate current reports in the light of our findings and give suggestions for future research.

The final section brings the debate back to the general issue of assessing low-probability risk. We stress that the findings in this paper are not to be interpreted as an argument for anti-intellectualism, but rather as arguments for making the noisy and fallible nature of scientific and technical research subject to intellectual reasoning, especially in situations where the probabilities are very low and the stakes very high.

**2. Probability Estimates**

Suppose you read a report which examines a potentially catastrophic risk and concludes that the probability of catastrophe is one in a billion. What probability should you assign to the catastrophe occurring? We argue that direct use of the report's estimate of one in a billion is naïve. This is because the report's authors are not infallible and their argument might have a hidden flaw. What the report has told us is not the probability of the catastrophe occurring, but the probability of the catastrophe occurring *given that the included argument is sound*. Even if the argument looks watertight, the chance that it contains a critical flaw may well be much larger than one in a billion. After all, in a sample of a billion apparently watertight arguments you are likely to see many that have hidden flaws. Our best estimate of the probability of catastrophe may thus end up noticeably higher than the report's estimate.[2]

Let us use the following notation:

---

[2] Scientific arguments are also sometimes erroneous due to deliberate fraud, however we shall not address this particular concern in this paper.

$X$ = the catastrophe occurs,

   $A$ = the argument is sound.

While we are actually interested in P($X$), the report provides us only with an estimate of P($X|A$), since it can't take into account the possibility that it is in error.[3] From the axioms of probability theory, we know that P($X$) is related to P($X|A$) by the following formula:

(1)    P($X$) = P($X|A$) P($A$) + P($X|\neg A$) P($\neg A$) .

To use this formula to derive P($X$) we would require estimates for the probability that the argument is sound, P($A$), and the probability of the catastrophe occurring given that the argument is unsound, P($X|\neg A$). We are highly unlikely to be able to acquire accurate values for these probabilities in practice but we shall see that even crude estimates are enough to change the way we look at certain risk calculations.

A special case, which occurs quite frequently, is for reports to claim that $X$ is completely impossible. However, this just tells us that $X$ is impossible given that all our current beliefs are correct, i.e. that P($X|A$) = 0. By equation (1) we can see that this is entirely consistent with P($X$) > 0, as the argument may be flawed.

Figure 1 is a simple graphical representation of our main point. The square on the left represents the space of probabilities as described in the scientific report, where the black area represents the catastrophe occurring and the white area represents it not occurring. The normalized vertical axis denotes the probabilities for the event occurring and not occurring. This representation ignores the possibility of the argument being unsound. To accommodate this possibility, we can revise it in the form of the square on the right. The black and white areas have shrunk in proportion to the probability that the argument is sound and a new grey area represents the possibility that the argument is unsound. Now the horizontal axis is also normalized and represents the probability that the argument is sound.

---

[3] An argument *can* take into account the possibility that a certain sub-argument is in error. For example, it could offer two alternative sub-arguments to prove the same point. We encourage such practice and look at an example in section 4. However, no argument can fully take into account the possibility that it is itself is flawed — this would require an additional higher-level argument.

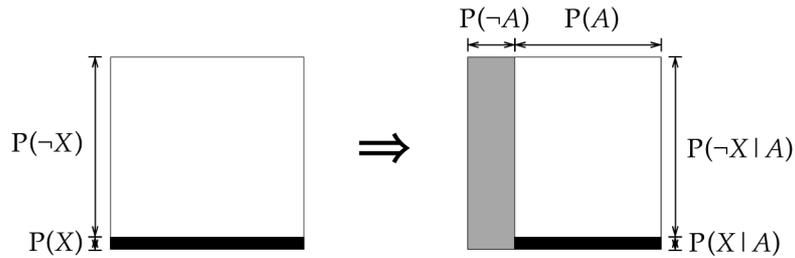

Figure 1: The left panel depicts a report's view on the probability of an event occurring. The black area represents the chance of the event occurring, the white area represents it not occurring. The right hand panel is the more accurate picture, taking into account the possibility that the argument is flawed and that we thus face an grey area containing an unknown amount of risk.

To continue our example, let us suppose that the argument made in the report looks very solid, and that our best estimate of the probability that it is flawed is one in a thousand, $(P(\neg A) = 10^{-3})$. The other unknown term in equation (1), $P(X|\neg A)$, is generally even more difficult to evaluate, but lets suppose that in the current example, we think it highly unlikely that the event will occur even if the argument is not sound, and that we also treat this probability as one in a thousand. Equation (1) tells us that the probability of catastrophe would then be just over one in a million — an estimate which is a thousand times higher than that in the report itself. This reflects the fact that if the catastrophe were to actually occur, it is much more likely that this was because there was a flaw in the report's argument than that a one in a billion event took place.

Flawed arguments are not rare. One way to estimate the frequency of major flaws in academic papers is to look at the proportion which are formally retracted after publication. While some retractions are due to misconduct, most are due to unintentional errors.[4] Using the MEDLINE database[5] (Cokol, Iossifov et al. 2007) found a raw retraction rate of $6.3 \cdot 10^{-5}$, but used a statistical model to estimate that the retraction rate would actually be between 0.001 and 0.01 if all journals received the same level of scrutiny as those in the top tier. This would suggest that $P(\neg A) > 0.001$, making our earlier estimate rather optimistic. We must also remember that an argument can easily be flawed without warranting retraction. Retraction is only called for when the underlying flaws are not trivial and are immediately noticeable by the academic community. The retraction rate for a field would thus provide a lower bound for the rate of serious flaws. Of course, we must also keep in mind the possibility that different branches of science may have different retraction rates and different error rates. In particular, the hard sciences may be less prone to error than the more applied sciences.

---

[4] Between 1982 and 2002, 62% of retractions were due to unintentional errors rather than misconduct (Nath, Marcus et al. 2006).

[5] A very extensive database of biomedical research articles from over 5,000 journals.

It is important to note the particular connection between the present analysis and high-stakes low-probability risks. While our analysis could be applied to any risk, it is much more useful for those in this category. For it is only when $P(X|A)$ is very low that the grey area has a relatively large role to play. If $P(X|A)$ is moderately high, then the small contribution of the error term is of little significance in the overall probability estimate, perhaps making the difference between 10% and 10.001% rather than the difference between 0.001% and 0.002%. The stakes must also be very high to warrant this additional analysis of the risk, for the adjustment to the estimated probability will typically be very small in absolute terms. While an additional one in a million chance of a billion deaths certainly warrants further consideration, an additional one in a million chance of a house fire does not.

One might object to our approach on the grounds that we have shown only that the uncertainty is greater than previously acknowledged, but not that the probability of the event is greater than estimated: the additional uncertainty could just as well decrease the probability of the event occurring. When applying our approach to arbitrary examples, this objection would succeed; however in this article, we are specifically looking at cases where there is an extremely low value of $P(X|A)$, so practically any value of $P(X|\neg A)$ will be higher and will thus drive the combined probability estimate upwards. The situation is symmetric with regard to extremely high estimates of $P(X|A)$, where increased uncertainty about the argument will reduce the probability estimate, the symmetry is broken only by our focus on arguments which claim that an event is very unlikely.

Another possible objection is that since there is always a nonzero probability of the argument being flawed, the situation is hopeless: any new argument will be unable to remove the grey area completely. It is true that the grey area can never be completely removed, however if a new argument ($A_2$) is independent of the previous argument ($A_1$) then the grey area will shrink, for $P(\neg A_1, \neg A_2) < P(\neg A_1)$. This can allow for significant progress. A small remaining grey area can be acceptable if $P(X|\neg A)P(\neg A)$ is estimated to be sufficiently small in comparison to the stakes.

### 3. Theories, Models and Calculations

The most common way to assess the reliability of an argument is to distinguish between model and parameter uncertainty and assign reliabilities to these choices. While this distinction has certainly been of use in many practical cases, it is unnecessarily crude for the present purpose, failing to account for potential errors in the paper's calculations or a failure of the background theory.

In order to account for all possible mistakes in the argument, we look separately at its *theory*, its *model*, and its *calculations*. The calculations evaluate a concrete model representing the processes under consideration, e.g. the formation of black holes in a particle collision, the response of certain climate parameters (such as mean temperature or precipitation rate) to changes in greenhouse gas concentrations, or the response of economies to changes in the oil price. These models are mostly derived from more general theories. In what follows, we do not restrict the term 'theory' to well-established and mathematically elaborate theories like electrodynamics, quantum chromodynamics or relativity theory. Rather, theories are understood to include theoretical background knowledge such as specific research paradigms or the generally accepted research practice within a field. An example is the efficient market hypothesis which underlies many models within economics, such as the Black-Scholes model.

Even incorrect theories and models can be useful, if their deviation from reality is small enough for the purpose at hand. Hence we consider *adequate* models or theories rather than *correct* ones. For example, we wish to allow that Newtonian mechanics is an adequate theory in many situations, while recognizing that in some cases it is clearly inadequate (such as for calculating the electron orbitals). We thus call a representation of some system adequate if it is able to predict the relevant system features at the required precision. For example, if climate modellers wish to determine the implications our greenhouse gas emissions will have on the well-being of future generations; their model/theory will not be adequate unless it tells them the changes in the *local* temperature and precipitation. In contrast, a model might only need to tell them changes in *global* temperature and precipitation to be adequate for answering less sensitive questions. On a theoretical level, much more could be said about this distinction between adequacy and correctness, but for the purposes of evaluating the reliability of risk assessment, the explanation above should suffice.

With the following notation:

$T$ = the involved theories are adequate

$M$ = the derived model is adequate

$C$ = the calculations are correct

we break down $A$ in the way indicated above and replace P($X|A$) in equation (1) by P($X|T,M,C$) and P($A$) by P($T,M,C$). From the laws of conditional probability it follows that:

(2)    P($T,M,C$) = P($T$) P($M|T$) P($C|M,T$)

We may assume $C$ to be independent of $M$ and $T$, as the correctness of a calculation is independent of whether the theoretical and model assumptions underpinning it were adequate. Given this independence, P($C|M,T$) = P($C$), so the above equation can be simplified:

(3)    P($T,M,C$) = P($T$) P($M|T$) P($C$).

Substituting this back into equation (1), we obtain a more tractable formula for the probability that the event in question occurs.

We have already made a rough attempt at estimating P($A$) from the paper retraction rates. Estimating P($T$), P($M|T$) and P($C$) is more accurate and somewhat easier, though still of significant difficulty. Though estimating the various terms in equation (3) must ultimately be done on a case by case basis, the following elucidation of what we mean by calculation, model and theory will shed some light on how to pursue such an analysis. By incorporating our threefold distinction, it is straightforward to apply findings on the reliability of theories from philosophy of science — based, for example, on probabilistic verification methods (e.g. (Reichenbach 1938)) or falsifications as in (Hempel 1950) or (Popper 1959). Often, however, the best we can do is to put some bounds upon them based on the historical record. We thus review typical sources of error in the three areas.

*3.1. Calculation: Analytic and Numeric*

Estimating the correctness of the calculation independently from the adequacy of the model and the theory seems important whenever the mathematics involved is non-trivial. Most cases where we are able to provide more than purely heuristic and hand-waving risk assessments are of this sort. Consider climate models evaluating runaway climate change and risk estimates for the LHC or for asteroid impacts. When calculations accumulate, even trivial mathematical procedures become error-prone. A particular difficulty arises due to the division of labour in the sciences: commonly in modern scientific practice, various steps in a calculation are done by different individuals who may be in different working groups in different countries. The Mars Climate Observer spacecraft was lost in 1999 because a piece of control software from Lockheed Martin used Imperial units instead of the metric units the interfacing NASA software expected (NASA 1999).

Calculation errors are distressingly common. There are no reliable statistics on the calculation errors made in risk assessment or, even more broadly, within scientific papers. However, there is research on errors made in some very simple calculations that performed in hospitals. Dosing errors give an approximate estimate of how often mathematical slips occur. Errors in drug charts occur at a rate of 1.2% to 31% across different studies (Prot, Fontan et al. 2005; Stubbs, Haw et al. 2006; Walsh, Landrigan et al. 2008), with a median of roughly 5% of administrations. Of these errors 15-40% were dose errors, giving an overall dose error rate of about 1–2%.

What does this mean for error rates in risk estimation? Since the stakes are high when it comes to dosing errors, this data represents a serious attempt to get the right answer in a life or death circumstance. It is likely that the people doing risk estimation are more reliable at arithmetic than health professionals and have more time for error correction, but it appears unlikely that they would be many orders of magnitude more reliable. Hence a chance of $10^{-3}$ for a mistake per simple calculation does not seem unreasonable. A random sample of papers from *Nature* and the *British Medical Journal* found that roughly 11% of statistical results were flawed, largely due to rounding and transcription errors (García-Berthou and Alcaraz 2004).

Calculation errors include more than just the 'simple' slips which we know from school, such as confusing units, forgetting a negative square root, or incorrectly transcribing from the line above. Instead, many mistakes arise here due to numerical implementation of the analytic mathematical equations. Computer based simulations and numerical analysis are rarely straightforward. The history of computers contains a large number of spectacular failures due to small mistakes in hardware or software. The June 4 1996 explosion of an Ariane 5 rocket was due to a leftover piece of code triggering a cascade of failures (ESA 1996). Audits of spreadsheets in real-world use find error rates on the order of 88% (Panko 1998). The 1993 Intel Pentium floating point error affected 3-5 million processors, reducing their numeric reliability and hence our confidence in anything calculated with them (Nicely 2008). Programming errors can remain dormant for a long time even in apparently correct code, only to emerge under extreme conditions. An elementary and widely used binary search algorithm included in the standard libraries for Java was found after nine years to contain a bug that emerges only when searching very large lists (Bloch 2006). A mistake in data-processing led to the retraction of five high-profile protein structure papers as the handedness of the molecules had become inverted (Miller 2006).

In cases where computational methods are used in modelling, many mistakes cannot be avoided. Discrete approximations of the often continuous model equations are used, and in some cases we know that the discrete version is not a good proxy for the continuous model (Morawetz and Walke 2003). Moreover, numerical evaluations are often done on a discrete computational grid, with the values inside the meshes being approximated from the values computed at the grid points. Though we know that certain extrapolation schemes are more reliable in some cases than others, we are often unable to exclude the possibility of error, or to even quantify it.

*3.2 Ways of modelling and theorizing*

Our distinction between model and theory follows the typical use of the terms within mathematical sciences like physics or economics. Whereas theories are associated with broad applicability and higher confidence in the correctness of their description, models are closer to the phenomena. For example, when estimating the probability of a particular asteroid colliding with the earth, one would use either Newtonian mechanics or general relativity as a theory for describing the role of gravity. One could then use this theory in conjunction with observations of the bodies' positions, velocities and masses to construct a model, and finally, one could perform a series of calculations based on this model to estimate the probability of impact. As this shows, the errors that can be introduced in settling for a specific model include and surpass those which are sometimes referred to as *parameter uncertainty*. As well as questions of the individual parameters (positions, velocities, masses), there are important questions of detail (can we neglect the inner structure of the involved bodies?), and breadth (can we focus on the Earth and asteroid only, or do we have to model other planets, or the Sun?).[6]

As can be seen from this example, one way to distinguish theories from models is that theories are too general to be applied directly to the problem. For any given theory, there are many ways to apply it to the problem and these ways give rise to different models. Philosophers of science will note that our theory/model distinction is in accordance with the non-uniform notion used by (Giere 1999), (Morrison 1998), (Cartwright 1999), and others, but differs from that of (Suppes 1957).

---

[6] This question of breadth is closely linked to what (Hansson 1996) refers to as *demarcation uncertainty*. But demarcation of the problem involves not only the obvious demarcation in physical space and time, but also questions of the systems to consider, the scales to consider etc.

We should also note that it is quite possible for an argument to involve several theories or several models. This complicates the analysis and typically provides additional ways for the argument to be flawed.[7] For example, in estimating the risk of black hole formation at the LHC, we not only require quantum chromodynamics (the theory the LHC is built to test), but also relativity and Hawking's theory of black hole radiation. In addition to their other roles, modelling assumptions also have to explain how to glue such different theories together (Hillerbrand and Ghil 2008).

In risk assessment, the systems involved are most often not as well understood as asteroid impacts. Often, various models exist simultaneously — all known to be incomplete or incorrect in some way, but difficult to improve upon. Particularly in these cases, having an expected or desired outcome in mind while setting up a model, makes one vulnerable to expectation bias: the tendency to reach the desired answer rather than the correct one. This bias has affected many of science's great names (Jeng 2006), and in the case of risk assessment, the desire for a 'positive' outcome (safety in the case of the advocate or danger in the case of the protestor) seems a likely cause of bias in modelling.

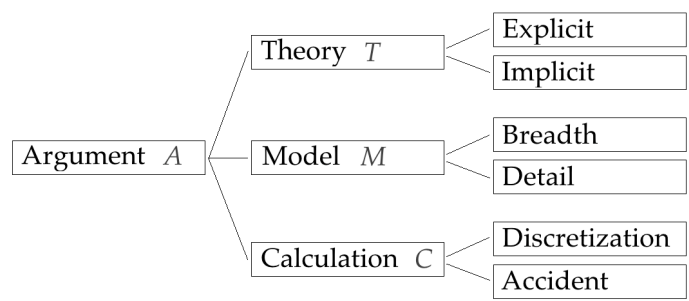

Figure 2: Our distinctions regarding the ways in which risk assessments can be flawed.

*3.3 Historical examples of Model and Theory Failure*

A dramatic example of a model failure was the Castle Bravo nuclear test on March 1 1954. The device achieved 15 megatons of yield instead of the predicted 4-8 megatons. Fallout affected parts of the Marshal Islands and irradiated a Japanese fishing boat so badly that one fisherman died, causing an international incident (Nuclear Weapon Archive 2006). Though the designers at Los Alamos National Laboratories understood the involved theory of alpha decay, their model of the reactions involved in the explosion was too narrow, for it neglected the decay of one of the involved particles (lithium-7), which turned out to contribute the bulk of the explosion's energy. The Castle Bravo test is also notable for being an example of model failure in a very serious experiment conducted in the hard sciences and with known high stakes.

---

[7] Additional theories and models can also be deliberately introduced in order to *lower* the probability of argument failure, and in section 4, we shall see how this has been done for the safety assessment of the LHC.

The history of science contains numerous examples of how generally accepted theories have been overturned by new evidence or understanding, as well as a plethora of minor theories that persisted for a surprising length of time before being disproven. Classic examples for the former include the Ptolemaic system, phlogiston theory and caloric theory; an example for the latter is human chromosome number, which was systematically miscounted as 48 (rather than 46) and this error persisted for more than 30 years (Gartler 2006).

As a final example, consider Lord Kelvin's estimates of the age of the Earth (Burchfield 1975). They were based on information about the earth's temperature and heat conduction, estimating an age of the Earth of between 20 and 40 million years. These estimates did not take into account radioactive heating, for radioactive decay was unknown at the time. Once it was shown to generate additional heat the models were quickly updated. While neglecting radioactivity today would count as a model failure, in Lord Kelvin's day it represented a largely unsuspected weakness in the physical understanding of the Earth and thus amounted to theory failure. This example makes it clear that the probabilities for the adequacy of model and theory are not independent of each other, and thus in the most general case we cannot further decompose equation (3).

**4. Applying our analysis to the risks from particle physics research**

Particle physics is the study of the elementary constituents of matter and radiation, and the interactions between them. A major experimental method in particle physics involves the use of particle accelerators such as the RHIC and LHC to bring beams of particles to near the speed of light and then collide them together. This focuses a large amount of energy in a very small region and breaks the particles down into their components, which are then detected. As particle accelerators have become larger, the energy densities achieved have become more extreme, prompting some concern about their safety. These safety concerns have focused on three possibilities: the formation of 'true vacuum', the transformation of the earth into 'strange matter', and the destruction of the earth through the creation of a black hole.

*4.1 True vacuum and strange matter formation*

The type of vacuum that exists in our universe might not be the lowest possible vacuum energy state. In this case, the vacuum could decay to the lowest energy state, either spontaneously, or if triggered by a sufficient disturbance. This would produce a bubble of 'true vacuum' expanding outwards at the speed of light, converting the universe into different state apparently inhospitable for any kind of life (Turner and Wilczek 1982).

Our ordinary matter is composed of electrons and two types of quarks: up quarks and down quarks. *Strange matter* also contains a third type of quark: the 'strange' quark. It has been hypothesized that strange matter might be more stable than normal matter, and able to convert atomic nuclei into more strange matter (Witten 1984). It has also been hypothesized that particle accelerators could produce small negatively charged clumps of strange matter, known as *strangelets*. If both these hypotheses were correct and the strangelet also had a high enough chance of interacting with normal matter, it would grow inside the Earth, attracting nuclei at an ever higher rate until the entire planet was converted to strange matter — destroying all life in the process. Unfortunately strange matter is complex and little understood, giving models with widely divergent predictions about its stability, charge and other properties (Jaffe, Busza et al. 2000).

One way of bounding the risk from these sources is the *cosmic ray argument*: the same kind of high-energy particle collisions occur all the time in Earth's atmosphere, on the surface on the Moon and elsewhere in the universe. The fact that the Moon or observable stars have not been destroyed despite a vast number of past collisions (many at much higher energies than can be achieved in human experiments) suggest that the threat is negligible. This argument was first used against the possibility of vacuum decay (Hut and Rees 1983) but is quite general.

An influential analysis of the risk from strange matter was carried out in (Dar, De Rujula et al. 1999) and formed a key part of the safety report for the RHIC. This analysis took into account the issue that any dangerous remnants from cosmic rays striking matter at rest would be moving at high relative velocity (and hence much less likely to interact) while head-on collisions in accelerators could produce remnants moving much at much slower speeds. They used the rate of collisions of cosmic rays in free space to estimate strangelet production. These strangelets would then be slowed by galactic magnetic fields and eventually be absorbed during star formation. When combined with estimates of the supernova rate, this can be used to bound the probability of producing a dangerous strangelet in a particle accelerator. The resulting probability estimate was $< 2 \cdot 10^{-9}$ per year of RHIC operation.[8]

While using empirical bounds and experimentally tested physics reduces the probability of a theory error, the paper needs around 30 steps to reach its conclusion. For example, even if there was just a $10^{-4}$ chance of a calculation or modelling error per step this would give a total $P(\neg A) \approx 0.3\%$. This would easily overshadow the risk estimate. Indeed, even if just one step had a $10^{-4}$ chance of error, this would overshadow the estimate.

A subtle complication in the cosmic ray argument was noted in (Tegmark and Bostrom 2005). The Earth's survival so far is not sufficient as evidence for safety, since we do not know if we live in a universe with 'safe' natural laws or a universe where planetary implosions or vacuum decay do occur but we have just been exceedingly lucky so far. While this latter possibility might sound very unlikely, all observers in such a universe would find themselves to be in the rare cases where their planets and stars had survived, and would thus have much the same evidence as we do. Tegmark and Bostrom had thus found that in ignoring these anthropic effects, the previous model had been overly narrow. They corrected for this anthropic bias and, using analysis from (Jaffe, Busza et al. 2000), concluded that the risk from accelerators was less than $10^{-12}$ per year.

---

[8] (Kent 2004) points out some mistakes in stating the risk probabilities in different versions of the paper, as well as for the Brookhaven report. Even if these are purely typesetting mistakes, it shows that the probability of erroneous risk estimates is nonzero.

This is an example of a demonstrated flaw in an important physics risk argument (one that was pivotal in the safety assessment of the RHIC). Moreover, it is significant that the RHIC had been running for five years on the strength of a flawed safety report, before Tegmark and Bostrom noticed and fixed this gap in the argument. Although this flaw was corrected immediately after being found, we should also note that the correction is dependent on both anthropic reasoning and on a complex model of the planetary formation rate (Lineweaver, Fenner et al. 2004). If either of these, or the basic Brookhaven analysis is flawed, the risk estimate is flawed.

*4.2 Black hole formation*

The Large Hadron Collider experiment at CERN was designed to explore the validity and limitations of the Standard Model of particle physics by colliding beams of high energy protons. This will be the most energetic particle collision experiment ever done, which has made it the focus of a recent flurry of concerns. Due to the perceived strength of the previous arguments on vacuum decay and strangelet production, most of the concern about the LHC has focused on black hole production.

None of the theory papers we have found appears to have considered the black holes to be a safety hazard, mainly because they all presuppose that any black holes would immediately evaporate due to Hawking radiation. However, it was suggested by (Dimopoulos and Landsberg 2001) that if black holes form, particle accelerators could be used to test the theory of Hawking radiation. Thus critics also began questioning whether we could simply assume that black holes would evaporate harmlessly.

A new risk analysis of LHC black-hole production (Giddings and Mangano 2008) provides a good example of how risks can be more effectively bounded through multiple sub-arguments. While never attempting to give a probability of disaster (rather concluding "there is no risk of any significance whatsoever from such black holes") it uses a multiple bounds argument. It first shows that rapid black hole decay is a robust consequence of several different physical theories ($A_1$). Second it discusses the likely incompatibility between non-evaporating black holes and mechanisms for neutralising black holes: in order for cosmic ray–produced stable black holes to be innocuous but accelerator-produced black holes to be dangerous, they have to be able to shed excess charge rapidly ($A_2$). Our current understanding of physics suggests both that black holes decay and that even if they didn't, they would be unable to discharge themselves. Only if this understanding is flawed will the next section come into play.

The third part, which is the bulk of the paper, models how multidimensional and ordinary black holes would interact with matter. This leads to the conclusion that if the size scale of multidimensional gravity is smaller than about 20 nm, then the time required for the black hole to consume the Earth would be larger than the natural lifetime of the planet. For scenarios where rapid Earth accretion is possible, the accretion time inside white dwarves and neutron stars would also be very short, yet production and capture of black holes from impinging cosmic rays would be so high that the lifespan of the stars would be far shorter than the observed lifespan (and would contradict white dwarf cooling rates) ($A_3$).

While each of these arguments have weaknesses the force of the total argument ($A_1, A_2, A_3$) is significantly stronger by the combination of them. Essentially the paper acts as three sequential arguments, each partly filling in the grey area (see figure 1) left by the previous. If the theories surrounding black hole decay fail, the argument about discharge comes into play, and if against all expectation black holes are stable and neutral the third argument shows that astrophysics constrains them to a low accretion rate.

*4.3 Implications for the safety of the LHC*

What are the implications of our analysis for the safety assessment of the LHC? First, let us consider the stakes in question. If one of the proposed disasters were to occur, it would mean the destruction of the earth. This would involve the complete destruction of the environment, 6.5 billion human deaths and the loss of all future generations. It is worth noting that this loss of all future generations (and with it, all of humanity's potential) may well be the greatest of the three, but a comprehensive assessment of these stakes is outside the scope of this paper. For the present purposes, it suffices to observe that the destruction of the earth is at *least* as bad as 6.5 billion human deaths.

There is some controversy as to how one should combine probabilities and stakes into an overall assessment of a risk. Some hold that the simple approach of expected utility is the best, while others hold some form of risk aversion. However, we can sidestep this dispute by noting that in either case, the risk of some harm is at least as bad as the expected loss. Thus, a risk with probability *p* of causing a loss at least as bad as 6.5 billion deaths is at least as bad as a certain $6.5 \cdot 10^9 p$ deaths.

Now let us turn to the best estimate we can make of the probability of one of the above disasters occurring during the operation of the LHC. While the arguments for the safety of the LHC are commendable for their thoroughness, they are not infallible. Although the report considered several possible physical theories, it is eminently possible that these are all inadequate representations of the underlying physical reality. It is also possible that the models of processes in the LHC or the astronomical processes appealed to in the cosmic ray argument are flawed in an important way. Finally, it is possible that there is a calculation error in the report.

Recall equation (1):

(1)   $P(X) = P(X | A) P(A) + P(X | \neg A) P(\neg A)$

P(X) is formed from two terms. The second of these represents the additional probability of disaster due to the argument being unsound. It is the product of the probability of argument failure and the probability of disaster given such a failure. Both terms are very difficult to estimate, but we can gain insight by showing the ranges they would have to lie within, for the risk presented by the LHC to be acceptable.

From (1), we obtain that:

(4)    $P(X) \geq P(X \mid \neg A) P(\neg A)$ .

If we let $l$ denote the acceptable limit of expected deaths from the operation of the LHC, we get: $6.5 \cdot 10^9 \, P(X) \leq l$. Combining this with equation (4), we obtain:

(5)    $P(X \mid \neg A) P(\neg A) \leq 1.5 \cdot 10^{-10} \, l$ .

This inequality puts a severe bound on the acceptable values for these probabilities. Since it is much easier to grasp this with an example, we shall provide some numbers for the purposes of illustration. Suppose, for example, that the limit were set at 1000 expected deaths, then $P(X \mid \neg A) P(\neg A)$ would have to be below $1.5 \cdot 10^{-7}$ for the risk to be worth bearing. This requires very low values for these probabilities. We have seen that for many arguments, $P(\neg A)$ is above $10^{-3}$. We have also seen that the argument for the safety of the RHIC turned out to have a significant flaw, which was unnoticed by the experts at the time. It would thus be very bold to suppose that the argument for the safety of the LHC was much lower than $10^{-3}$, but for the sake of argument, let us grant that it is as low as $10^{-4}$ — that out of a sample of 10,000 independent arguments of similar apparent merit, only one would have any serious error.

Even with the value of $P(\neg A)$ were as low as $10^{-4}$, $P(X \mid \neg A)$ would have to be below 0.15% for the risk to be worth taking. $P(X \mid \neg A)$ is the probability of disaster given that the arguments of the safety report are flawed and is the most difficult component of equation (1) to estimate. Indeed, few would dispute that we really have very little idea of what value to put on $P(X \mid \neg A)$. It would thus seem overly bold to set this below 0.15% without some substantive argument. Perhaps such an argument could be provided, but until it is, such a low value for $P(X \mid \neg A)$ seems unwarranted.

We stress that the above combination of numbers was purely for illustrative purposes, but we cannot find any plausible combination of the three numbers which meets the bound and which would not require significant argument to explain either the levels of confidence or the disregard for expected deaths. We would also like to stress that we are open to the possibility that additional supporting arguments and independent verification of the models and calculations could significantly reduce the current chance of a flaw in the argument.

However, our analysis implies that the current safety report should not be the final word in the safety assessment of the LHC. To proceed with the LHC on the arguments of the most recent safety report alone, we would require further work on estimating $P(\neg A)$, $P(X \mid \neg A)$, the acceptable expected death toll, and the value of future generations and other life on earth. Such work would require expertise beyond theoretical physics, and an interdisciplinary group would be essential. If the stakes were lower, then it might make sense for pragmatic concerns to sweep aside this extra level of risk analysis, but the stakes are astronomically large, and so further analysis is critical. Even if the LHC goes ahead without any further analysis, as is very likely, these lessons must be applied to the assessment of other high-stakes low-probability risks.

**5. Conclusions**

When estimating threat probabilities, it is not enough to make conservative estimates (using the most extreme values or model assumptions compatible with known data). Rather, we need robust estimates that can handle theory, model and calculation errors. The need for this becomes considerably more pronounced for low-probability high-stake events, though we do not say that low probabilities cannot be treated systematically. Indeed, as pointed out by (Yudkowsky 2008), if we could not correctly predict probabilities lower than $10^{-6}$, we could not run lotteries.

Some people have raised the concern that our argument might be too powerful: for it is impossible to disprove the risk of even something as trivial as dropping a pencil, then our argument might amount to prohibiting everything. It is true that we cannot *completely* rule out any probability that apparently inconsequential actions might have disastrous effects, but there are a number of reasons why we do not need to worry about universal prohibition. A major reason is that for events like the dropping of a pencil which have no plausible mechanism for destroying the world, it seems just as likely that the world would be destroyed by *not* dropping the pencil. The expected losses would thus balance out. It is also worth noting that our argument is simply an appeal to a weak form of decision theory to address an unusual concern: for our method to lead to incorrect conclusions, it would require a flaw in decision theory itself, which would be very big news.

It will have occurred to some readers that our argument is fully applicable to this very paper: there is a chance that we have made an error in our own arguments. We entirely agree, but note that this possibility does not change our conclusions very much. Suppose, very pessimistically, that there is a 90% chance that our argument is sufficiently flawed that the correct approach is to take safety reports' probability estimates at face value. Even then, our argument would make a large difference to how we treat such values. Recall the example from section 2, where a report concludes a probability of $10^{-9}$ and we revise this to $10^{-6}$. If there is even a 10% chance that we are correct in doing so, then the overall probability estimate would be revised to $0.9 \cdot 10^{-9} + 0.1 \cdot 10^{-6} \approx 10^{-7}$, which is still a very significant change from the report's own estimate. In short, even serious doubt about our methods should not move one's probability estimates more than an order of magnitude away from those our method produces. More modest doubts would have a effect.

The basic message of our paper is that any scientific risk assessment is only able to give us the probability of a hazard occurring conditioned on the correctness of its main argument. The need to evaluate the reliability of the given argument in order to adequately address the risk was shown to be of particular relevance in low-probability high-stake events. We drew a three-fold distinction between theory, model and calculation, and showed how this can be more useful than the common dichotomy in risk assessment between model and parameter uncertainties. By providing historic examples for errors in the three fields, we clarified the three-fold distinction and showed where flaws in a risk assessment might occur. Our analysis was applied to the recent assessment of risks that might arise from experiments within particle physics. To conclude this paper, we now provide some very general remarks on how to avoid argument flaws when assessing risks with high stakes.

Firstly, the testability of predictions can help discern flawed arguments. If a risk estimate produces a probability distribution for smaller, more common disasters this can be used to judge whether the observed incidences are compatible with the theory. Secondly, reproducibility appears to be the most effective way of removing many of these errors. By having other people replicate the results of calculations independently our confidence in them can be dramatically increased. By having other theories and models *independently* predict the same risk probability our confidence in them can again be increased, as even if one of the arguments is wrong the others will remain. Finally, we can reduce the possibility of unconscious bias in risk assessment through the simple expedient of splitting the assessment into a 'blue' team of experts attempting to make an objective risk assessment and a 'red' team of devil's advocates attempting to demonstrate a risk, followed by repeated turns of mutual criticism and updates of the models and estimates (Calogero 2000). Application of such methods could in many cases reduce the probability of error by several orders of magnitude.